\documentclass{svjour2}                    
\smartqed  
\usepackage{graphicx,amssymb,epsfig}
\newcommand{\be}{\begin{equation}}
\newcommand{\ee}{\end{equation}}
\newcommand{\beq}{\begin{eqnarray}}
\newcommand{\eeq}{\end{eqnarray}}
\newcommand{\beqn}{\begin{eqnarray*}}
\newcommand{\eeqn}{\end{eqnarray*}}
\def\op{ \ $ }
\def\cl{$ \ }
\def\lappreq{\mathrel{\rlap{\lower2.5pt\hbox{\hskip1pt$\sim$}}
    \raise1pt\hbox{$<$}}}         
\def\gappreq{\mathrel{\rlap{\lower2.5pt\hbox{\hskip1pt$\sim$}}
    \raise1pt\hbox{$>$}}}         
\def\apra{{\rm APR1~}}
\def\aprb{{\rm APR2~}}
\def\apba{{\rm APRB200~}}
\def\apbb{{\rm APRB120~}}
\def\BBSA{{\rm BBS1~}}
\def\BBSB{{\rm BBS2~}}
\def\GG{{\rm G240~}}
\def\lsim{\mathrel{\rlap{\lower2.5pt\hbox{\hskip1pt$\sim$}}
    \raise1pt\hbox{$<$}}}         
\def\gsim{\mathrel{\rlap{\lower2.5pt\hbox{\hskip1pt$\sim$}}
    \raise1pt\hbox{$>$}}}         
\def\msun{M_\odot}
\begin{document}
\title{Quasi-Normal Modes and Gravitational Wave Astronomy}
\titlerunning{Quasi-Normal Modes and Gravitational Wave Astronomy}
\authorrunning{V. Ferrari, L. Gualtieri}
\author{Valeria Ferrari \and Leonardo Gualtieri}
\institute{ Dipartimento di Fisica ``G. Marconi'', Sapienza Universit\`a 
di Roma and INFN, Sezione Roma 1, P.le A. Moro
2, 00185 Roma, Italy}
\maketitle
\begin{abstract}
We review the main results obtained in the literature
on quasi-normal modes of compact stars and black holes, in the 
light of recent exciting developments of gravitational wave detectors.
Quasi-normal modes are a fundamental feature of the
gravitational signal emitted  by compact objects in many astrophysical processes; 
we will show that their eigenfrequencies
encode interesting information on the nature and on the inner structure of the
emitting source and we will  discuss  whether we are ready for a gravitational wave
asteroseismology.
\end{abstract}
\section{Introduction}
In the past four years the sensitivity of the 
gravitational wave (GW) detectors LIGO and Virgo has been improved
at a formidable rate \cite{webaddress}.  LIGO's noise curve has been lowered
by about three orders of magnitude and now the detectors are operating
at  the design sensitivity.  Similar progresses have been achieved by Virgo, although
some effort remains to be done to reach the planned sensitivity at low
frequencies ($\sim (10-40)$~Hz).  In any event, these detectors are now
in a position to take data good enough to start making science: a
supernova exploding in the local group of nearby galaxies would not be
missed, as well as the coalescence of compact bodies, neutron stars or
black holes, with total mass smaller than about 100 $M_\odot$, out to a
distance of the orders of few Megaparsecs (these estimates are only
indicative, since they are continuously updated as the detector
sensitivities are improved).  The detection of  gravitational signals 
will allow to test the predictions of the theoretical work that 
has been done over the years to construct waveforms and energy spectra, 
and to extract distinctive features which could be traced back to the
nature and to the structure of the source. 
An important piece of information
is provided by the frequencies at which a compact object
oscillates and emits gravitational waves, i.e the quasi-normal mode
(QNM) frequencies.  In this paper we shall discuss the pulsation properties 
of  black holes and neutron stars, focussing  in particular on
the information they carry about the emitting source.

Accounts on quasi-normal modes of stars and black holes 
can also be found in \cite{Kostasrev} and \cite{Nollertrev}.
\section{Do black holes  oscillate?}
According to General Relativity,  quasi-normal modes are the proper
modes at which a black hole, or a star, oscillates when excited by a
non radial perturbation.  They are said {\it quasi-normal}, 
in contrast to the {\it normal} modes of Newtonian gravity,
because they  are damped by the emission of gravitational waves; as a consequence,
the  corresponding eigenfrequencies are complex.
That  a star can oscillate is, in some sense, obvious because a star is a ball of
fluid \footnote{ The study of stellar oscillations started at the
beginning of the past  century, when Shapley \cite{sha} (1914) and
Eddington \cite{edd1} (1918) suggested that the variability observed in
some stars is due to periodic pulsations.}; however, when the idea that also
black holes possess some proper modes of vibration  was firstly proposed, 
it raised considerable
surprise. Indeed,  a black hole is not a material object, it is a singularity
hidden by a horizon: how can it possibly oscillate?  In order to
understand how this bizarre behaviour was discovered, we need to go
back to half a century ago and to the early theory of black hole
perturbations.


\subsection{Schwarzschild perturbations are described by two wave equations}
In 1957 T. Regge and J.A. Wheeler \cite{regwe}  showed that the
equations describing the perturbations of a Schwarzschild black hole
can be separated if the perturbed metric tensor is expanded in
tensorial spherical harmonics.  They also showed that the relevant
equations split into two decoupled sets belonging to different parity
--- \op (-1)^\ell\cl ({\it even} or {\it polar}) and \op (-1)^{\ell+1}\cl
({\it odd} or {\it axial}) --- and that, by a suitable choice of the
gauge and by Fourier-expanding the perturbed functions, 
the equations for the radial part of the axial perturbations of
a Schwarzschild black hole can be reduced to a single
Schroedinger-like wave equation with a potential barrier, for a
suitably defined function $Z_\ell^-$:
\beq
\label{RZ}
\frac{d^{2}Z_\ell^-}{dr_{*}^{2}}+[\omega^{2}-V_\ell^-(r)]Z_\ell^-=0,
\eeq
where
\beq
\label{potaxial}
V_\ell^-(r)=\frac{1}{r^{3}}\left( 1-\frac{2M}{r}\right) [\ell(\ell+1)r-6M]\,,
~~
r_*=r+2M\log (\frac{r}{2M}-1).
\eeq
This equation is known as the Regge-Wheeler equation. A similar result
was obtained later in 1970 by F. Zerilli \cite{zer}, who showed that
also the polar equations can be reduced to the wave equation
(\ref{RZ}) for a suitably defined function $Z_\ell^+$, and with a
different potential barrier
\beq
\label{potz}
V_\ell^+(r)=\frac{2(r-2M)} {r^{4}(nr+3M)^{2}}
[n^{2}(n+1)r^{3}+3Mn^{2}r^{2}+ 9M^{2}nr+9M^{3}]\; ,
\eeq
where $n=\frac{1}{2}(\ell-1)(\ell +2)$.
If the perturbation is excited by a source, on the left-hand
side there will be a forcing term obtained from the harmonic expansion
of the stress-energy tensor of the exciting source.  The wave
equations for the axial and polar perturbations describe the way in
which a non rotating black hole reacts to an external perturbation,
and the gravitational signal emitted by
the perturbed black hole can be calculated in terms of the two
functions $Z_\ell^-$ and $Z_\ell^+$ as follows:
\begin{eqnarray}
h^+(t,r,\theta,\phi)&=&\frac{1}{2\pi}\int
 \frac{e^{{\rm i}\omega(t-r_*)}}{r}\sum_{\ell m}\left[
Z^+_{\ell m}(r,\omega)W^{\ell m}(\theta,\phi)
-\frac{Z^-_{\ell m}(r,\omega)}{{\rm i}\omega}
\frac{X^{\ell m}(\theta,\phi)}{\sin\theta}\right]d\omega\nonumber\\
h^\times(t,r,\theta,\phi)&=&\frac{1}{2\pi}\int
\frac{e^{{\rm i}\omega(t-r_*)}}{r}
\sum_{\ell m}\left[Z^+_{\ell m}(r,\omega)\frac{X^{lm}(\theta,\phi)}{\sin\theta}+
\frac{Z^-_{\ell m}(r,\omega)}{{\rm i}\omega}
W^{\ell m}(\theta,\phi)\right]d\omega
\nonumber\\
\end{eqnarray}
where $h^+$, $h^\times$ are the two polarizations of the gravitational
wave in the transverse-traceless gauge (see \cite{MTW}, Chapter 35),
and
\begin{eqnarray}
W^{\ell m}(\theta,\phi)&=&\left(\partial_\theta^2-\cot\theta\partial_\theta
-\frac{1}{\sin^2\theta}\partial_\phi^2\right)Y^{\ell m}(\theta,\phi)
\nonumber\\
X^{\ell m}(\theta,\phi)&=&2\left(\partial_{\theta\phi}-\cot\theta\partial_\phi
\right)Y^{\ell m}(\theta,\phi)
\end{eqnarray}
where $Y^{\ell m}$ are  scalar spherical harmonics.
Note that, since we are considering perturbations of a spherically symmetric
spacetime, $Z^\pm_{\ell m}$  coincides with $Z^\pm_{\ell}$ for any value of $m$.

\subsection{Quasi-normal modes of a Schwarzschild black hole}
In 1970 Vishveshwara \cite{vish} pointed out that  equation
(\ref{RZ}) for the functions $Z_\ell^-$ and $Z_\ell^+$ allows complex
frequency solutions which satisfy the following boundary conditions
\beqn
&&Z_\ell \rightarrow e^{i \omega r_*},\qquad r_* \rightarrow -\infty,\\
&&Z_\ell \rightarrow e^{-i \omega r_*},\qquad r_* \rightarrow +\infty;\\
\eeqn
the former represents a pure ingoing wave,
since nothing can escape from a black
hole horizon, the latter represents a pure outgoing wave at radial infinity and
corresponds to the requirement that no radiation is incoming from infinity.
This idea was confirmed by Press \cite{pr1} who found, by integrating
the wave equation numerically, that an arbitrary initial perturbation
decays as a pure frequency mode. However, only in 1975 Chandrasekhar
and Detweiler \cite{chdet} actually computed the discrete
eigenfrequencies of these modes and clarified their nature.  Quoting
Chandrasekhar from its book {\it The Mathematical Theory of Black
Holes} \cite{mt}: 
\par\noindent``.. we may expect on general grounds that any initial
perturbation will, during its last stages, decay in a manner
characteristic of the black hole and independently of the original
cause. In other words, we may expect that during the very last stages,
the black hole will emit gravitational waves with frequencies and
rates of damping, characteristic of itself, in the manner of a bell
sounding its last dying pure note. These considerations underlie the
formulation of the concept of the {\it quasi-normal modes} of a black
hole.''

A Schwarzschild black hole is characterized by only one
parameter, its mass $M$; consequently, the QNM frequencies depend only on $M$.
In Table \ref{table1} we show the 
values of the complex characteristic frequencies of the first four
QNMs of a Schwarzschild black hole, respectively for
\op\ell=2\cl and \op\ell=3.\cl
\begin{table}[ht]
\caption{The lowest QNM  frequencies of a Schwarzschild black hole for 
\op\ell=2\cl and \op\ell=3.\cl
They are the same both for the polar and for the axial perturbations, i.e.  {\it the two
potential barriers (\ref{potaxial}) and (\ref{potz}) are isospectral.}
}
\vspace{0.4cm}
\begin{center}
\begin{tabular}{|c|c|c|c|}
\hline
&${M\omega_0+iM\omega_i}$&&${M\omega_0+iM\omega_i}$ \\
\hline
$\ell=2$&  0.3737+i0.0890&
$\ell=3$&  0.5994+i0.0927 \\
 \hline
& 0.3467+i0.2739&& 0.5826+i0.2813 \\
\hline
& 0.3011+i0.4783&& 0.5517+i0.4791 \\ \hline
& 0.2515+i0.7051&& 0.5120+i0.6903 \\ \hline
\end{tabular}
\end{center}
\label{table1}
\end{table}
In order to find the true pulsation frequency, $\nu$,
 and the damping time, $\tau$, from the values given
in Table \ref{table1}, we proceed as follows.
Let us assume that the black hole mass is \op M=n
M_\odot,\cl ($M_\odot=1.48\cdot 10^5$~ cm); converting to physical
unities we find 
\be
\nu=\frac{c}{2\pi n\cdot M_\odot (M\omega_0)}
=\frac{32.26}{n}(M\omega_0) ~{\rm kHz},\quad
\tau=\frac{n M_\odot}{ (M\omega_i)c}=\frac{n\cdot 0.4937\cdot
10^{-5}}{ (M\omega_i)} ~{\rm s}.
\ee
Using these expressions we can check whether a gravitational signal
emitted by an oscillating black hole falls within the bandwidth of the
ground based interferometers Virgo/LIGO or within that of the space
based interferometer LISA.  Virgo/LIGO bandwidth extends over a range
of frequencies which goes from about 10-40 Hz, up to few kHz.  Thus,
these detectors will be able to detect the signal emitted by an oscillating black hole
(if it is sufficiently strong) with mass ranging within
\[
10~ M_\odot \lappreq M \lappreq 10^3~ M_\odot,
\]
corresponding to the frequency range $ \nu \in$ [12 Hz, 1.2 kHz];
LISA will be sensitive to the frequency region  $ \nu \in
[10^{-4},10^{-1}]$ Hz, and will see oscillating black holes with mass
\[
1.2\cdot 10^5 ~ M_\odot \lappreq M \lappreq 1.2\cdot 10^8~ M_\odot.
\]
For instance, LISA will be able to detect signals emitted by the oscillations of the
massive black hole at the center of our Galaxy SGR A*, the mass of which is
$M=(3.7\pm 0.2)\cdot 10^6~M_\odot$ \cite{Ghez:2003qj}.

While the frequencies of the lowest modes are rather easy to compute,
great care must be used to determine the entire spectrum. 
Many different methods have been used to this purpose.
For instance, a WKB approximation and a higher order WKB approach
have been used to find  the lowest \cite{wkb1} and the higher mode  frequencies
\cite{wkb2}, respectively.  
In addition, new approaches have been developed to study the QNM spectrum, as
the continued fraction method (developed for Kerr black holes in \cite{leaver})
and the phase-integral method \cite{phase}. 
Using these approaches, it has been found  \cite{Asint} that,
for any value of the harmonic index $\ell$,
as the order $n$ of the mode increases, the real part of
a mode frequency approaches a non zero limiting value.
Furthermore, an analytical expression has been found for the imaginary 
part of the  frequency, valid in the limit $n\rightarrow\infty$.
We also mention that an exact, analytical solution of the
Regge-Wheeler equation has recently been found in terms of the Heun functions.
QNM eigenfrequencies can be computed in terms
of this solution, by solving numerically a boundary value problem
\cite{Fiziev}.

\subsection{The quasi-normal modes of a Kerr black hole}
After 1975  the study of black hole perturbations follows along
two principal avenues. One studies directly the
perturbations of the metric tensor via Einstein's equations
linearized about a given background. The other studies the
perturbations of Weyl's and Ricci's scalars using
the Newman-Penrose formalism.  Using this latter approach in 1972
Teukolsky \cite{teuk} was able to decouple and separate
the equations governing the perturbations of a Kerr black hole, and to
reduce them to a single master equation for the radial part of the
perturbation $ R_{\ell m}$: 
\beq
\label{teukolski}
\cases{
\Delta R_{\ell m,rr}+2(s+1)(r-M)R_{\ell m,r}+V(r)R_{\ell m}=0 &\cr
V_{\ell m}(r,\omega)=
\frac{1}{\Delta}\left[ (r^2+a^2)^2\omega^2-4aMrm\omega+a^2m^2
+2is(am(r-M)\right.&\cr
\left.-M\omega(r^2-a^2))\right]r
+\left[ 2is\omega r-a^2\omega^2-A_{\ell m}\right]&\cr
\Delta =r^2-2Mr+a^2. &\cr
}
\eeq
The angular part, $ S_{\ell m},$ satisfies the equations of the oblate
spheroidal harmonics
\beq
\label{angular}
\cases{
[(1-u^2)S_{\ell m,u}]_{,u}+\left[
a^2\omega^2u^2-2am\omega su+s+A_{\ell m}-\frac{(m+su)^2}{1-u^2}
\right]S_{\ell m}=0,&\cr
 u=\cos\theta, &\cr
}
\eeq
and the complete wavefunction is
\beq
\psi_s(t,r,\theta,\varphi)=\frac{1}{2\pi}
\int e^{{\rm i}\omega t}\sum_{\ell=\vert
s\vert}^\infty\sum_{m=-\ell}^\ell e^{im\varphi}S_{\ell m}(u)R_{\ell m}(r)d\omega.
\eeq
In these equations   $s$ is the spin-weight parameter  
$ s=0,\pm 1,\pm 2,$ for scalar, electromagnetic and
gravitational perturbations, respectively
\footnote{$+$ and $-$ indicate the ingoing and outgoing radiative part of
the considered field. For example \op s=+2\cl corresponds to \op
\psi_2=\Psi_0,\cl and \op s=-2\cl to \op \psi_{-2}=\Psi_4,\cl where
\op \Psi_0\cl and \op \Psi_4\cl are the Weyl scalars.}, and 
$A_{\ell m}$ is a separation constant.  

It is interesting to note  that,
unlike the potential barrier of a Schwarzschild black hole, which is
real and independent of the harmonic index $m$ and of the frequency, the potential barrier 
of a Kerr black hole is complex, and it  depends  on $m$ and on the frequency
$\omega$.

An interesting phenomenon occurs when
electromagnetic or gravitational waves are scattered by Kerr's
potential barrier $V_{\ell m}(r,\omega)$; if the incident wave has a
frequency in the range
\beq
0 < \omega < \omega_c \qquad\hbox{where}\qquad \omega_c
=\frac{am}{2Mr_+},\quad m>0,
\eeq
the reflection coefficient associated to $V(r)$ exceeds
unity \cite{starchu,prteu}. This phenomenon is called {\it
superradiance}, and it is the analogue, in the domain of wave
propagation, of Penrose's process in the domain of particle
creation.

The quasi-normal frequencies of a Kerr black hole have been 
computed by Detweiler \cite{detkerr}, and subsequently by Leaver
\cite{leaver}, Seidel \& Iyer \cite{seiy}, Kokkotas \cite{kk1} and
Onozawa \cite{recentKerrmodes}.  Since rotation removes the
degeneracy presented by Schwarzschild's modes, there is
a set of eigenmodes for any assigned value of the
harmonic indexes $\ell$ and $m$.
The calculations show that when the black hole angular momentum  $a$ increases,
the real part of the complex eigenfrequencies
is bounded, but the imaginary part is
not. Moreover, when a Kerr black hole becomes ``extreme'', i.e. when
\op a\rightarrow M,\cl  highly damped
mode-frequencies converge to the purely real value of the critical
frequency below which superradiant scattering occurs, \op\omega_c=\frac{m}{2M}.\cl In
this context, an interesting result was obtained by Detweiler in 1977
\cite{detkerr}. He found that when 
\op a\rightarrow M,\cl the imaginary part of the
mode-frequencies tends to zero. If excited, these modes would set
the black hole into an oscillation that would never decay, suggesting
that extreme Kerr black holes are ``marginally unstable''.  It was
subsequently shown by B. Mashoon and one of the authors 
that when \op
a\rightarrow M\cl the amplitude of the ``unstable'' modes tends to zero, and
consequently quasi-normal modes belonging to  real frequency cannot exist in the ordinary
regime \cite{mashoonferrari}.

As for a  Schwarzschild black hole, 
in the $n\rightarrow\infty$ limit the imaginary part of the mode frequencies has
an analytical expression which has been determined in
\cite{AsintKerr}.
\subsection{Excitation of black hole quasi-normal modes}
\begin{figure}[ht]
\begin{center}
\includegraphics[width=7cm,angle=270]{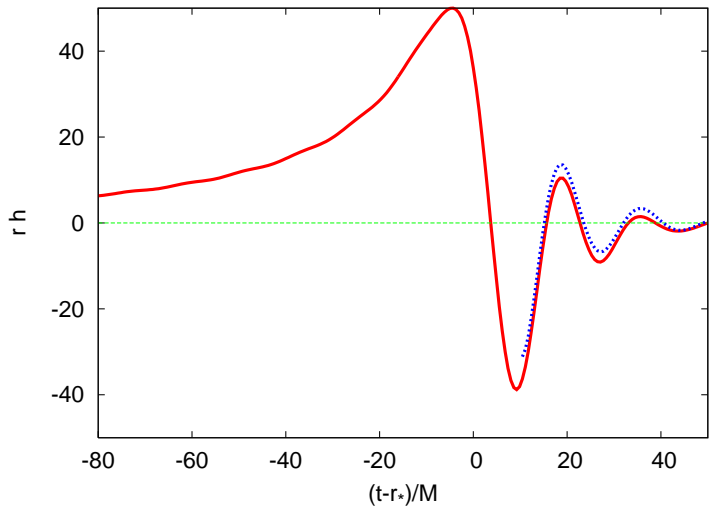}
\caption{The gravitational signal emitted when a Schwarzschild black
hole is perturbed by a radially infalling particle  (solid
line);  the analytical fit (dashed line) has been obtained using a linear combination of
the first two $\ell=2$ black hole quasi-normal modes.}
\label{tail}
\end{center}
\end{figure}

Black holes quasi-normal modes  are excited in many astrophysical
processes, and are a fundamental feature of the gravitational
signal.  Therefore they are  of utmost importance for the data
analysis of gravitational wave experiments.
The first simulation of black hole oscillations excited by an
external source dates back to 1971: in  \cite{DRPP}
a Schwarzschild black hole was perturbed by a radially infalling
point-like body, with a mass much smaller than the black hole mass.
The energy spectrum of the emitted signal 
was computed by solving the Zerilli equation with
a source describing the infalling particle. The waveform 
was  explicitly computed in \cite{RTV}, and it was shown 
that, after a transient, the signal exhibits a {\it ringing tail},
which can be fitted by a combination of quasi-normal
modes.  In Figure \ref{tail}, we show the gravitational wave amplitude
$r\cdot h(t-r_*)$ emitted in the considered process,
and the analytical fit with 
the first two  $\ell=2$ quasi-normal modes belonging to the frequencies
$M\omega_1=0.37+{\rm i} 0.09$ and $M\omega_2=0.35+{\rm i} 0.27$.
The fit becomes more
accurate if higher order modes are taken into account, but the main
contribution is due to $\omega_1$ and $\omega_2$.

The perturbation induced on a  Schwarzschild black hole by extended sources made
up of pressureless matter was studied in \cite{SW}. It was shown
that in this case modes are also excited, but to a smaller extent 
with respect to the  pointlike source, due to interference effects
(see also \cite{ON,BCW}).
In the case of matter orbiting around a black hole, the excitation of
the QNMs is in general even smaller. Indeed, modes are significantly
excited only if matter reaches $r\lesssim 4M$, i.e., if it crosses
the potential barrier $V^\pm_\ell(r)$; however, this is
not possible if  matter moves on a stable orbit, for which
$r\ge 6M$ \cite{NOK}.

Although these studies refer to very idealised situations,
they have been very useful because they showed that 
quasi-normal modes can be excited, and because they provided
a first  understanding of the mechanisms underlying the mode excitation.
However, astrophysical phenomena are much more complicate, and only recently
major advances in numerical techniques allowed the
modelling of more realistic processes involving black holes.

Black hole coalescence is probably the most violent
process occurring in the universe (after the big bang), and it is
expected to be the most powerful source of gravitational waves to be
detected by interferometric detectors
Virgo and LIGO.
\vskip .2cm
\begin{figure}[ht]
\begin{center}
\includegraphics[width=7cm]{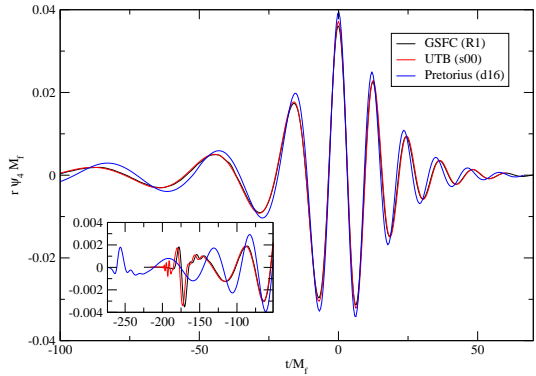}
\caption{Gravitational radiation waveform emitted by two coalescing 
black holes (courtesy of the authors).}
\label{compcoal}
\end{center}
\end{figure}
Such phenomena are very difficult to simulate numerically,
because the topology of the spacetime changes
during the process. After nearly 15 years of efforts, this problem has
been solved very recently by three groups \cite{numcoal}, who found,
independently, the same results \cite{comparcoal}. In their
simulations, two Schwarzschild black holes coalesce, merge and produce
a single Kerr black hole.
The emitted gravitational signal computed in \cite{comparcoal} is shown in Figure
\ref{compcoal}:   after a short merger
phase the waveform is clearly dominated by the quasi-normal mode oscillations.
This important feature is also exhibited  by the signal emitted
in the coalescence of rotating black holes \cite{coalkerr}.

Quasi-normal modes excitation has also been shown to give a strong contribution
to the signal emitted after the core collapse of a rotating
neutron star to a Kerr black hole  \cite{BHRS}.
Moreover,  in \cite{ZFR} the gravitational emission of a black hole perturbed by a thick,
oscillating accretion disk has been  studied; in a subsequent paper
\cite{FGR} it has been shown  that
if the disk is extremely  dense, black hole oscillations can be excited, even though,
due to the symmetry of the source, only by a small amount.
All numerical simulations of astrophysical processes in which QNMs are  excited  show that
the leading contribution belong to the lowest
frequency, $\ell=2$ mode.

To conclude this Section, we should at least mention that  there are also
theoretical studies \cite{leaver2},\cite{residues} 
on the ``excitability'' of the quasi-normal modes. 
The main result of these studies is that a measure of
the relative QNM excitation, independent of the particular
astrophysical process, is encoded in the poles of the Green's
functions associated to the Zerilli and Regge-Wheeler equations.  This
information, combined with the knowledge of the exciting source, allows
to determine the quasi-normal mode content of the gravitational
signal.
\subsection{Other issues on quasi-normal modes}
\subsubsection{Completeness}
The quasi-normal modes of a Schwarzschild black hole 
do not form a complete basis for
black hole perturbations.  As shown by Leaver
\cite{leaver2}, this is due to a branch cut in the Green's functions associated to
the Zerilli and Regge-Wheeler equations. Consequently, a general 
perturbation cannot be written as a combination of quasi-normal modes. In
particular, it exhibit a  power-law tail
$\sim t^{-2\ell -2}$ \cite{price72}.

\subsubsection{Stability}
The stability of the Schwarzschild spacetime has been proved by
Vishveshwara \cite{Vishstab} and Wald \cite{Waldstab}.  Vishveshvara
showed that the imaginary part, $\omega_i$, of the QNM frequency
is always positive, and Wald proved that since  $\omega_i$ is always
positive,  all perturbations remain bounded.

The stability of Kerr black holes is still an open issue. The main problem arises
because of the existence of the ergoregion, where  a perturbation can grow
indefinitely even though the energy remains finite. However, there are
indications that also  Kerr black holes are stable \cite{KerrStability}.

\subsubsection{Applications in string theory and in loop quantum gravity}
In recent years, it has been suggested that black hole  QNM's 
may play a role in  string theory and loop quantum gravity.
In 2000 Horowitz and Hubeny \cite{HH} proposed that the study of
the  black hole QNM's in anti-de Sitter spacetime could be useful to
determine some properties of conformal field theories. Their conjecture
is deeply rooted in string theory and in the so-called ``AdS-CFT
correspondence''. Stimulated by this work, many authors computed
the QNM eigenfrequencies  in anti-de Sitter spacetime \cite{AdS} (see
also \cite{AdSold}).  
It is worth reminding that  anti-de Sitter solution of Einstein's 
equations describes a universe with a negative cosmological constant;
therefore these black holes should not be considered as astrophysical objects.

In 2003 Dreyer and Motl \cite{DM} suggested that, 
in the asymptotic limit $n\rightarrow\infty$,
black hole quasi-normal modes would allow 
to fix the value of the ``Immirzi parameter'', which is a
key parameter in loop quantum gravity. Following this proposal,
studies of the asymptotic limit of QNM \cite{Asint}, \cite{AsintKerr}
have further been developed \cite{Asint2}.

More generally, inspired by these consideration in the contexts of
string theory and loop quantum gravity, in recent years many authors
have computed the eigenfrequencies of black hole quasi-normal modes
in various background spacetimes,
both in four dimensions and for higher dimensional spacetimes
\cite{VariousQNM}.


\section{Stellar pulsations}
Stellar pulsations  are a very well known phenomenon in astronomy, since
they  underlay a variety of astrophysical processes.
For instance, they are observed in the Sun,
and  a branch of solar sciences, named {\it
helioseismology},  uses the information encoded in the pulsation frequencies
to investigate the internal structure of our star and the physical processes 
that occur in the interior.
Non radial pulsations  are associated to gravitational wave emission
and, as we shall see, the mode frequencies carry interesting information on
the  inner structure of the emitting sources. Thus, if in the future
GW-detectors will be able to catch the gravitational signals emitted by  pulsating
stars, a new branch of astrophysics will develop, the 
{\it gravitational wave asteroseismology}.
This  will allow us to investigate  the interior of neutron stars,
where densities and pressures are so extreme that they are unreachable by 
high energy experiments on Earth. But before discussing how the 
equation of state (EOS) of matter  affects 
quasi-normal mode frequencies,
let us preliminarily show the equations we need
to solve to determine these frequencies.
We shall discuss only pulsations of a non rotating star,
i.e.  of  stars which are described by  static, spherically symmetric
solutions of Einstein's equations.
The rotating case is much more complicate, and
an exhaustive description of the problems that arise when one is
looking for the quasi-normal mode frequencies is beyond the scope of this
paper (see, for instance, \cite{Nick} and references therein).

\subsection{Stellar perturbations of a non rotating star}
The equations governing the adiabatic
 perturbations of a spherical star in general
relativity have been derived within different approaches by many
authors \cite{TC}-\cite{DL}. 
Here we shall show, as an example,
 the basic equations of the theory of stellar
perturbations as developed in \cite{CF} (see also \cite{VF} for a complete
account of the theory).
We start with the metric appropriate to describe a static, spherical background 
\be
ds^{2}=e^{2\nu}(dt)^{2}-e^{2\psi}d\varphi
-e^{2\mu_{2}}(dr)^{2}-e^{2\mu_{3}}(d\theta)^{2}.
\label{spa}
\ee
where $e^{2\psi}=r^2$, $e^{2\mu_{3}}=r^2\sin^2\theta$, and $\nu(r)$ and $\mu_2(r)$ have to be
found by solving the TOV equations of stellar structure 
(see for instance \cite{MTW}),
for an assigned equation of state.
Matter in the star is assumed to be a perfect fluid, with stress-energy tensor
\[
T^{\mu\nu}= (\epsilon+p)u^\mu u^\nu - p g^{\mu\nu},
\]
where $\epsilon(r)$ and $p(r)$ are the fluid energy-density and the pressure,
 and $u^\mu$ are the components of the four-velocity of a generic fluid element.
Axisymmetric perturbations of the spacetime (\ref{spa})
can be described by the line-element
\be
\label{nonstazionaria}
ds^{2}=e^{2\nu}(dt)^{2}-e^{2\psi}(d\varphi
-q_{2}dr^{2}-q_{3}d\theta-\omega
dt)^{2}-e^{2\mu_{2}}(dr)^{2}-e^{2\mu_{3}}(d\theta)^{2}.
\ee
In eq. (\ref{nonstazionaria})
there are seven  unknown functions, i.e. one more than
needed, but the extra degree of freedom disappears when the
boundary conditions of the problem are fixed.
As a consequence of a generic perturbation, the metric functions,
$(\nu,\psi,\mu_2,\mu_3,\omega,q_2,q_3)$,
and the fluid variables, $(\epsilon,p,u^\alpha)$, 
change by a small amount with respect to their
unperturbed values, which we assume to be known; for instance
$\nu  \longrightarrow \nu+\delta\nu,$
$\epsilon \longrightarrow\epsilon+\delta\epsilon$, 
and similarly for the remaining variables.
At the same time, each element of fluid  undergoes
an infinitesimal displacement from its equilibrium position,
which is described by the lagrangian displacement $\vec \xi.$
All perturbed quantities are functions of  $t, r$ and $\theta$.
If we now write Einstein's equations supplemented by
the hydrodynamical equations and the conservation of baryon number,
expand all tensors in tensorial spherical harmonics and
Fourier-expand the time dependent quantities,
we find that, as for black holes, the perturbed equations
decouple into two sets,
the {\it polar} and the {\it axial}, with a major
difference: the {\it polar} equations 
couple the thermodynamical variables to the metric variables.
Conversely the {\it axial} perturbations 
do not induce fluid motion except for a stationary rotation;
however, we shall see that the fluid plays a role,
because it shapes the potential barrier associated to the spacetime
curvature.

\subsubsection{The equations for the polar perturbations}
The explicit expressions of the functions that describe the polar
perturbations, expanded in harmonics and Fourier-expanded are
\beq
&\delta \nu =N_\ell(r)P_{\ell}(\cos\theta)e^{i\omega t}&
\delta\mu_2=L_\ell(r)P_{\ell}(\cos\theta )e^{i\omega t}\\
\nonumber
&\delta\mu_3=[ T_\ell(r)P_{\ell}+V_\ell(r)P_{\ell,\theta,\theta }]e^{i\omega t}&
\delta\psi =[ T_\ell(r)P_{\ell}+
V_\ell(r)P_{\ell,\theta }\cot\theta]e^{i\omega t} ,\\
\nonumber
&\delta p=\Pi_\ell(r)P_{\ell}(\cos\theta)e^{i\omega t}
&2(\epsilon +p)e^{\nu+\mu_{2}}\xi_{r}(r,\theta)e^{i\omega t}
=U_\ell(r)P_{\ell}e^{i\omega t}   \\
\nonumber
&\delta\epsilon =E_\ell(r)P_{\ell}(\cos\theta)e^{i\omega t}
&2(\epsilon
+p)e^{\nu+\mu_{3}}\xi_{\theta}(r,\theta)e^{i\omega t}
=W_\ell(r)P_{\ell,\theta}e^{i\omega t},
\eeq
where $P_\ell(\cos\theta)$ are Legendre's polynomials and $\omega$ is the frequency.
After separating the variables the relevant Einstein's
equations become
\be
\cases{
X_{\ell,r,r}+\left(\frac{2}{ r}+\nu_{,r}-\mu_{2,r}\right)
X_{\ell,r}+\frac{n}{ r^{2}}
e^{2\mu_{2}}(N_\ell+L_\ell)+\omega^{2}e^{2(\mu_{2}-\nu)}X_\ell=0,
&\cr
(r^{2}G_\ell)_{,r}=n\nu_{,r}(N_\ell-L_\ell)+
\frac{n}{ r}(e^{2\mu_{2}}-1)(N_\ell+L_\ell)
+r(\nu_{,r}-\mu_{2,r})X_{\ell,r}
+\omega^{2}e^{2(\mu_{2}-\nu)}rX_\ell,
&\cr
-\nu_{,r}N_{\ell,r}=
- G_\ell+\nu_{,r}[X_{\ell,r}+\nu_{,r}(N_\ell-L_\ell)]
+\frac{1}{ r^{2}}(e^{2\mu_{2}}-1)
(N_\ell-rX_{\ell,r}-r^{2}G_\ell)&\cr
- e^{2\mu_{2}}(\epsilon+p)N_\ell
+\frac{1}{ 2}\omega^{2}e^{2(\mu_{2}-\nu)}
\left\{ N_\ell+L_\ell+\frac{r^{2}}{ n}G_\ell+
\frac{1}{ n}[rX_{\ell ,r}+(2n+1)X_\ell ]\right\} ,&\cr
L_{\ell,r}(1-{D})+ L_\ell\left[ \left(
\frac{2}{ r}-\nu_{,r}\right)-\left(\frac{1}{ r}
+\nu_{,r}\right)D\right]+
X_{\ell,r}+X_\ell\left(\frac{1}{ r}-\nu_{,r}\right)+
{D}N_{\ell,r}+&\cr
+N_\ell \left( {D}\nu_{,r}
-\frac{{D}}{ r}-{F}\right)+
\left(\frac{1}{ r}+{E}\nu_{,r}\right)\left[
N_\ell-L_\ell+\frac{r^{2}}{ n}G_\ell+
\frac{1}{ n}\left(rX_{\ell,r}+X_\ell\right)\right]=0,&\cr }
\label{compl}
\ee
where
\be
\nonumber
\cases{
{A}=\frac{1}{ 2}\omega^{2}e^{-2\nu},\qquad\quad
Q=\frac{(\epsilon +p)}{\gamma p},&\cr
\gamma=\frac{(\epsilon +p)}{p}\big (\frac{\partial p}{\partial\epsilon }
\big )_{entropy=const},\qquad\quad
{B}=\frac{e^{-2\mu_{2}}\nu_{,r}}{ 2(\epsilon+p)}
(\epsilon_{,r}-Qp_{,r}),&\cr
{D}=1-\frac{{A}}{ 2({A}+{B})}=
1-\frac{\omega^{2}e^{-2\nu}(\epsilon+p)}{
\omega^{2}e^{-2\nu}(\epsilon+p)+
e^{-2\mu_{2}}\nu_{,r}(\epsilon_{,r}-
Q p_{,r})},&\cr
{E}={D}(Q-1)-Q,&\cr
{F}=\frac{\epsilon_{,r}-Qp_{,r}}{ 2({A}+{B})}
=\frac{2\left[\epsilon_{,r}-Qp_{,r}\right](\epsilon+p)}
{2\omega^{2}e^{-2\nu}(\epsilon+p)+
e^{-2\mu_{2}}\nu_{,r}
(\epsilon_{,r}-Q p_{,r})},
&\cr}
\label{entireset}
\ee
and \op V_\ell\cl and \op T_\ell\cl have been replaced by
\op X_\ell\cl and \op G_\ell\cl  defined as
\be
\nonumber
\cases{
X_\ell=nV_\ell&\cr
\nonumber
G_\ell=\nu_{,r}[\frac{n+1}{ n}X_\ell-T_\ell]_{,r}+
\frac{1}{ r^{2}}(e^{2\mu_{2}}-1)
[n(N_\ell+T_\ell)+N_\ell]&\cr
+\frac{\nu_{,r}}{ r}(N_\ell+L_\ell)
-e^{2\mu_{2}}(\epsilon+p)N_\ell+
\frac{1}{ 2}\omega^{2}
e^{2(\mu_{2}-\nu)}[L_\ell-T_\ell+\frac{2n+1}{ n}X_\ell].&\cr}
\ee
Equations  (\ref{compl}) are valid in general,
also for non-barotropic equations of state.

It should be noted that although eqs. (\ref{compl})
describe stellar perturbations inside the star, they are written for
the variables $(X,G,N,L)$ which are metric perturbations.
However, the motion of
the fluid is excited by the perturbation and it can be shown that,
once equations (\ref{compl}) have been solved, 
the fluid variables, ($\Pi,E,U,W$), can be obtained in terms of the
metric functions using  the following equations
\beqn
&& W_\ell =T_\ell-V_\ell+L_\ell ,
\\
&&\Pi_\ell  = -\frac{1}{2}\omega^{2}e^{-2\nu}W_\ell-
(\epsilon +p)N_\ell ,\qquad
E_\ell  =  Q\Pi_\ell+
\frac{e^{-2\mu_{2}}}{2(\epsilon +p)}
(\epsilon_{,r}-Qp_{,r})U_\ell ,
\\
&&U_\ell =\frac{ [(\omega^{2}e^{-2\nu}W_\ell)_{,r}+
(Q+1)\nu_{,r}(\omega^{2}e^{-2\nu}W_\ell)+2(\epsilon_{,r}
-Qp_{,r})N_\ell ](\epsilon+p)}{
\left[ \omega^{2}e^{-2\nu}(\epsilon+p)+
e^{-2\mu_{2}}\nu_{,r}(\epsilon_{,r}-
Qp_{,r})\right]} .
\eeqn
This fact is remarkable: it shows that
all information on the dynamical behaviour
of a star is encoded in the gravitational field.
Thus, if one is interested exclusively in the study of the
emitted gravitational radiation, one can solve the system
(\ref{compl}) disregarding the fluid variables
\footnote{After these equations were derived, R.Ipser and R.H.Price
showed that they can be reduced to a fourth-order system \cite{ipserprice}.}.

Equations  (\ref{compl}) have to be
integrated for assigned values of the frequency from
\op r=0,\cl where  all functions must be regular,
 up to the stellar surface.
There, the spacetime becomes  vacuum and spherically
symmetric, and the perturbed metric functions and
their first derivatives have to be  matched continuously
with the Zerilli function that describes
the polar perturbations of a Schwarzschild spacetime; its expression in terms
of the metric functions is
\be
Z^{+}_\ell(r)=
\frac{r}{nr+3M}\left( {3M}X_\ell(r)/n-rL_\ell(r)\right)
\label{zerif}
\ee
(for a detailed discussion of the boundary conditions see
refs. \cite{CF} and \cite{VF}).

\subsubsection{A Schroedinger equation for the axial perturbations}
The equations for the axial perturbations are much simpler
than the polar ones.  Their radial behaviour is  completely described by
a function $ Z^{-}_\ell(r)$,
which satisfies the following Schroedinger-like equation
\be
\frac{d^{2}Z^{-}_\ell}{ dr_{*}^{2}}+
[\omega^{2}-V^{-}_\ell(r)]Z^{-}_\ell=0,
\label{axial}
\ee
where
\op
r_*=\int_0^r e^{-\nu+\mu_2}dr
\cl ($\nu$ and $\mu_2$ are unperturbed metric functions), and
\be
\label{zz1}
V^{-}_{\ell}(r)=\frac{e^{2\nu(r)}}{ r^{3}}\left\{\ell(\ell+1)r+r^{3}
\left[\epsilon(r) -p(r)\right]-6m(r)\right\}.
\ee
The function $Z^{-}_\ell(r)$ is a combination of the axial, metric perturbations
\be
\label{xdef}
e^{3\psi+\nu-\mu_2-\mu_3}\left(
\delta q_{2,\theta}-\delta q_{3,r}\right)=rZ_\ell^-(r) 
C^{-\frac{3}{2}}_{\ell+2}(\theta),
\ee
and $C^{-\frac{3}{2}}_{\ell+2}(\theta)$ are Gegenbauer's polynomials \cite{CF}.

Outside the star $\epsilon$ and $p$ vanish  and $V^{-}_{\ell}(r)$
reduces to  the Regge-Wheeler potential barrier (\ref{potaxial}).
From  these equations we see that
unlike the polar perturbations, the axial perturbations, which do not have a Newtonian
counterpart, are not coupled to fluid motion.
In addition, the shape of the potential (\ref{zz1}) depends on $\epsilon(r)$ and $p(r)$,
i.e. on the radial profile of  the energy-density
and of the pressure inside the star, in the equilibrium configuration.

\subsection{Quasi-normal modes of stars}
The quasi-normal modes  are solutions of
the axial and polar equations that satisfy
the following boundary conditions.
As for black holes, at radial infinity the solution 
must behave as a pure outgoing wave 
\be
Z^{\pm}_\ell\quad\rightarrow\quad e^{-i\omega t},
\qquad r_* \rightarrow +\infty.\\
\label{outgoing1}
\ee
In addition, all perturbed functions must be regular  at \op r=0\cl
and have to  match continuously  the
exterior perturbation on the stellar surface.
For the axial perturbations the matching condition is automatically satisfied,
because eq. (\ref{axial}) reduces to the Regge-Wheeler equation for $r \ge
R$, where $R$ is the stellar radius.

Stars possess many different classes of modes.
The axial quasi-normal modes are pure spacetime modes and do not exist in
Newtonian gravity.  They are named $w$-modes and are highly damped,
i.e.  the imaginary part of the frequency
is comparable to the real part \cite{wkostas}
 and consequently the damping times are small.
If the star is extremely compact, the potential (\ref{zz1}) inside the star
becomes a well, while in the exterior it remains a barrier. 
If the well is deep
enough, it allows for the existence of one or more slowly damped quasi-normal
modes, or $s$-modes; they are also said {\it trapped modes}
because, due to the slow damping, 
they are effectively trapped by the potential barrier,
and no much radiation can leak out of the star when these modes are
excited \cite{axialmodes}.

It is interesting to compare the eigenfrequencies of the axial modes of
stars and black holes, since they are both pure spacetime modes.
As an example, in table 2 we show the frequencies and the damping 
times of the first four \op\ell=2,\cl axial  modes for a
homogeneous  star with mass $ M=1.35 M_\odot$ and increasing compactness, and
for a non rotating black hole with the same mass.
\begin{table}[ht]
\caption{The characteristic frequencies and damping times of the axial
quasi-normal modes of a homogeneous star of mass $M= 1.35~\msun$.
The data are tabulated for  increasing values of the stellar compactness
${M\over R}$; they  are compared to those of a non rotating 
black hole with the same mass.
We tabulate the first four values of the frequency (in kHz) and of the damping time
(in s) for $\ell=2$. $\nu^s$ and $\tau^s$ refer to the {\it trapped}
modes associated to the potential well inside the star (see text),
$\nu^w$ and $\tau^w$ refer to the axial $w$-modes, and $\nu^{BH}$ and
$\tau^{BH}$
to the black hole}
\begin{center}
\renewcommand{\arraystretch}{1.4}
\setlength\tabcolsep{5pt}
\begin{tabular}{lllllll}
\hline\noalign{\smallskip}
${M\over R}$&$\nu^s$& $\tau^s$&$\nu^w$& $\tau^w$& $\nu^{BH}$&
$\tau^{BH}$\\
\noalign{\smallskip}
\hline
0.4167& 8.6293& $1.52\cdot 10^{-3}$ &11.1738 &
$1.70\cdot 10^{-4}$&8.9300&$7.49\cdot 10^{-5}$ \\
& -- & --  & 14.2757 &$8.03\cdot 10^{-5}$&8.2848&$2.43\cdot 10^{-5}$
\\
\\
& -- & -- & 18.2232 &$5.70\cdot 10^{-5}$& 7.1952& $1.39\cdot 10^{-5}$
\\
& -- & -- & 22.6669  &$4.88\cdot 10^{-5}$&
6.0099 &$0.95\cdot 10^{-5}$
\\
\hline
0.4386&4.4333 & 10.8 &10.4128  &$5.45\cdot 10^{-4}$&&
\\
&6.0168&  $2.50\cdot 10^{-1}$&11.9074  &$2.91\cdot 10^{-4}$&&
\\
&7.5462 &$ 1.44\cdot 10^{-2}$  & 13.4813 &$2.07\cdot 10^{-4}$&&
\\
&8.9891 &$ 1.83\cdot 10^{-3}$ &15.1428 &$1.67\cdot 10^{-4}$&&
\\
\hline
0.4425&2.6041 &$5.38\cdot 10^{3}$  & 10.7852  &$7.60\cdot 10^{-4}$&&
\\
&3.5427& $1.69\cdot 10^{2}$ &11.6922 &$5.34\cdot 10^{-4}$&&
\\&4.4802 &$1.22\cdot 10^{1}$   & 12.6138 &$4.22\cdot 10^{-4}$&&
\\
&5.4127 & $1.37\cdot 10^{-1}$ &13.5512 &$3.56\cdot 10^{-4}$&&
\\
\hline
\end{tabular}
\end{center}
\label{table2}
\end{table}
It should be mentioned that the $w$-mode frequencies depend upon the
equation of state of matter in the inner core of the star and, as shown in \cite{omarema},
if detected they would allow to discriminate between the models underlying 
different equations of state.
Until very recently,  the common belief was that
$w$-modes are unlikely to be  excited in astrophysical
processes. However in 2005 it has been shown that, in the collapse of a neutron star to a black
hole, $w$-modes are excited soon before the black hole forms. Thus, 
the gravitational signal emitted in this process
contains both the frequency of the $w$-modes of the collapsing star, 
and those of the quasi-normal modes of the newly born black hole \cite{BHRS}.

The polar modes are classified following 
a  scheme,  introduced by Cowling  in Newtonian gravity in 1942
\cite{cowling}, based on the restoring force which 
prevails when the generic fluid element is displaced from the equilibrium
position.  They are said  $g$-modes, or gravity modes, 
if the restoring force is due to buoyancy, 
and $p$-modes if it is due to pressure gradients.
The mode frequencies are  ordered as follows
\[
..\omega_{g_n} < .. < \omega_{g_1} <
\omega_f < \omega_{p_1}< .. < \omega_{p_n}..
\]
and are separated by the frequency of the fundamental mode ($f$-mode),
which has an intermediate character between $g-$ and $p-$
modes.
In addition, general relativity predicts the existence of
polar $w$-modes, that are very weakly coupled to
fluid motion and are similar to the axial $w$-modes \cite{kokkoschutzwmodes}.
Their frequencies are typically higher than those of the fluid modes ($g$, $f$ and $p$).

The relevance of the different modes to gravitational wave emission
depends on several factors, first of all
on the amount of energy which can be stored into a given mode.
Moreover, it depends  on the presence of other dissipative processes that may compete 
with GW-emission in removing energy from the star; 
for instance, neutrino diffusion or viscosity, if the
oscillating star is a newly born, hot neutron star.
Last, but not least, of crucial importance are
the values of the mode frequencies: for instance the
$w$-modes of neutron stars have typical frequencies of the order of
$10$ kHz or higher, far too high to be  detectable by ground based
interferometers in their present or advanced configuration.

Numerical simulations of the most energetic astrophysical processes, 
like core collapse to a neutron star (NS) or binary coalescence 
leading to NS formation, indicate that the mode  which is most 
excited  is the fundamental mode.  
For this reason  we will now
focus on this mode,  discussing the information the $f$-mode frequency,
$\nu_f$, carries  on the inner structure of  neutron stars.  

It is worth mentioning that  typical values of  $f$-mode damping times 
are of the order of a few tenths of seconds; 
consequently, the excitation of the $f$-mode would appear,
in the Fourier transform of the gravitational signal,
 as a sharp peak. Therefore, the $f$-mode frequency could, in principle,
be extracted from the detector noise by an appropriate data analysis
(provided $\nu_f$ falls in the detection bandwidth of some GW-detector).

From Newtonian gravity we know  that
the $\nu_f$ scales with the average density of the star
$ \nu_{f} \sim  \left(\frac{M}{R^3}\right)^{1/2}$;  
as shown in \cite{AK,astero_nostro}, where $\nu_f$  has been
computed for a variety of equations of state proposed to describe matter
in a NS, a similar scaling law holds also in general relativity. 

At densities exceeding  the equilibrium density of nuclear matter,
$\rho_0 =2.67 \times 10^{14}$ g/cm$^3$, the fluid in the inner core of a NS
is basically a gas of interacting nucleons. The equations of state proposed in the
literature to describe this kind of matter
are derived within two main, different approaches: 
nonrelativistic nuclear many-body theory (NMBT) and 
relativistic mean field theory (RMFT);   we will now show how 
different ways of modeling  hadronic interactions affect
the pulsation properties of the star.
In what follows we shall summarize the main results of a study
we have done in \cite{astero_nostro}, where
we have selected a restricted number of EOS
obtained within the NMBT and the RMFT approach.
To describe the inner and outer crust of the NS, we have  used
the Baym-Pethick-Sutherland
EOS \cite{BPS} and the Pethick-Ravenhall-Lorenz  EOS \cite{PRL},
respectively. 

The EOS we choose to describe NS matter at $\rho > \rho_0$,
are the following.\\
For the NMBT approach we select two groups of EOS:
{\bf Group I}, named  (\apra, \aprb, \apba, \apbb), and
{\bf Group II}, named (\BBSA, \BBSB), respectively.
In both cases matter is composed of neutrons, protons, electrons and muons in weak
equilibrium, and the dynamics is described by a non-relativistic Hamiltonian
which includes phenomenological potentials that describe 
two- and three-nucleon interactions.  The potential are obtained from fits 
of existing scattering data. 
For all EOS the two-body potential is $v_{18}$, whereas the
three-body potential is Urbana IX for {\bf Group I}, and Urbana VII for 
{\bf Group II}.

A first major difference between the two groups is that
in {\bf Group I} the ground state energy is
calculated using variational techniques \cite{AP,APR},
whereas in {\bf Group II} is calculated using
G-matrix perturbation theory \cite{BBS200}.
There are also differences among the EOS in each group:
\begin{itemize}
\item{} {\bf Group I} \\
\aprb is  an improved version of the \apra model.
In \apra  nucleon-nucleon potentials
describe interactions between nucleons in
their center of mass frame, in which the total momentum ${\bf P}$
vanishes.  In the \aprb  the two-nucleon  potential is
modified including relativistic corrections which arise from the boost to
a frame in which ${\bf P} \neq 0$, up to order ${\bf P}^2/m^2$.  These
corrections are necessary to use the nucleon-nucleon potential in a
locally inertial frame associated to the star. 
As a consequence of this change, the three-body
potential also needs to be modified in a consistent
fashion. 

The EOS \apba  and \apbb are the same as \aprb 
up to $\sim 4 \rho_0$, but at higher density there is 
a  phase of deconfined quark matter described
within the MIT bag model.  The mass of the strange quark is assumed to be
$m_s = 150$ MeV, the coupling constant describing quarks interaction is set
to $\alpha_s = 0.5$, and  the value of the bag
constant is 200 MeV/fm$^3$ for \apba and 120 MeV/fm$^3$ for \apbb.
We will discuss in some more detail quark matter in the next
section. The phase transition from nuclear matter to quark matter
is described requiring the fulfillment of Gibbs conditions, 
leading to the formation of a mixed
phase, and neglecting surface and Coulomb effects \cite{AP,BR}.
Thus, these stars are {\it hybrid} stars.

\item{} {\bf Group II} \\
The main difference between the equations of state \BBSA and \BBSB
is that in \BBSB strange heavy baryons
($\Sigma^-$ and $\Lambda^0$) are allowed to form in the core.
Neither \BBSA nor \BBSB include relativistic corrections.
\end{itemize}
As representative of the RMFT, we choose the EOS named \GG .
Matter composition includes leptons and the complete octet of
baryons (nucleons, $\Sigma^{0,\pm}$, $\Lambda^0$ and
$\Xi^\pm$). Hadron dynamics is described in terms of exchange of one
scalar and two vector mesons.  It should be reminded that in this case,
the EOS is obtained within the mean field approximation \cite{Gbook}.

\begin{figure}[h]
\begin{center}
\includegraphics[width=7cm,angle=270]{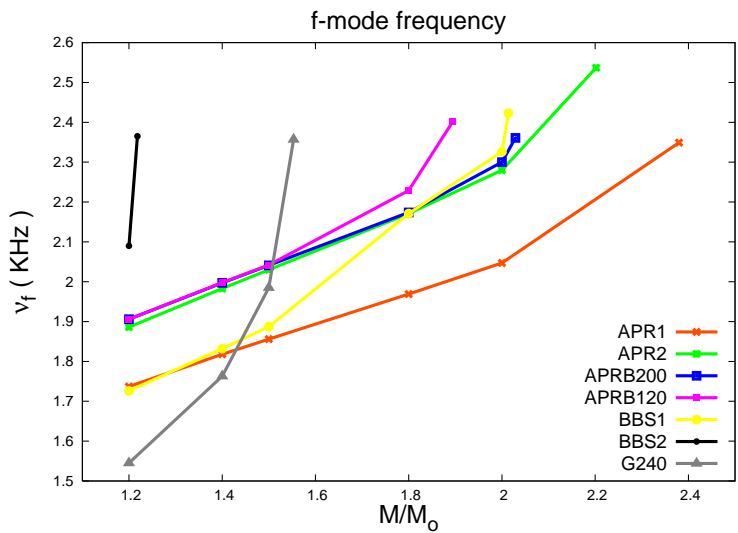}
\caption{The frequency  of the fundamental mode is plotted as a function of
the mass of the star for the selected EOS (see text).}
\label{nufM}
\end{center}
\end{figure}

For any of the above EOS 
we have solved the TOV equations for
different values of the central density,
finding the equilibrium  configurations. Then,
for each EOS and for each equilibrium model,  we have solved the
equations of stellar perturbations  finding the $f$-mode
frequency, $\nu_f$.
The results are shown  in Fig. \ref{nufM} where we plot 
$\nu_f$ as a function of the
mass, up to the maximum mass allowed by each EOS.
From this picture we learn the following.
Comparing the values of $\nu_f$ for \apra and \aprb we immediately see
that the relativistic corrections and the associated redefinition of
the three-body potential, which improve the Hamiltonian of \aprb with
respect to \apra, play a relevant role, leading to a systematic
difference of about 150 Hz in the mode frequency.  Conversely, the
presence of quark matter in the star inner core (EOS \apba and
\apbb) does not seem to significantly affect the pulsation 
properties of the star. 
We also see that the frequencies corresponding to
the \BBSA and \apra models, which are very close at $M \lsim 1.4 ~M_\odot$,
diverge for larger masses. This behavior can be traced back to the
different treatments of three-nucleon interactions, whose role in
shaping the EOS becomes more and more important as the star mass (and
central density) increases: while the variational approach of
ref. \cite{AP} used to derive the EOS \apra naturally allows for
inclusion of the three-nucleon potential appearing in
the Hamiltonian,  in G-matrix perturbation theory used to derive
the EOS \BBSA the three-body potential has to be replaced with an effective
two-nucleon potential, obtained by averaging
over the position of the third particle \cite{lejeune}.

The transition to hyperonic matter, predicted by the \BBSB model,
produces a considerable softening of the EOS, and leads to stable NS
configurations of very low mass ($=1.218~M_\odot$). 
As a consequence of the softening of
the EOS, the average density increases and so does the 
$f$-mode frequency, which is significantly higher
than that obtained for other EOS  for the same mass. 

It is also interesting to compare the $f$-mode frequencies corresponding
to models \BBSB (derived within the RMFT approach) and \GG
(derived using the NMBT approach), as they both predict the occurrence of heavy
strange baryons, but are obtained from different theoretical
approaches. 
The behavior of $\nu_f$ shown in Fig. \ref{nufM}
directly reflects the relations between mass and central density;
indeed, for a given mass,  larger central density
correspond to  smaller radii, and therefore  to larger average density.
Consequently, we can say that higher frequencies  correspond to
larger central  densities $\rho_c$. For example, the NS configurations of mass
1.2 $\msun$ correspond to $\rho_c\sim 7\cdot 10^{14}~$
g/cm$^3$   for  \GG,  and to a larger central density,
$\rho_c \sim 2\cdot 10^{15}~$g/cm$^3$ for \BBSB.
On the other hand, the
\GG model requires a central density of $\sim$ 2.5$\cdot$10$^{15}~$
g/cm$^3$ to reach a mass of $\sim$ 1.55 $\msun$ and a value of
$\nu_f$ equal to that of the \BBSB model.

From the above discussion we see that the frequency of the fundamental mode
carries interesting information on the different 
ways of modeling hadronic interactions. 
 
\subsection{Can a star  be made of quarks?}
Another interesting possibility is that a compact star is a strange star,
namely one that, except for a thin outer crust, is
entirely made of  a degenerate gas of up, down and strange 
quarks. That such stars may exist, was suggested  by Witten many years ago
\cite{Witten};  since then, from time to time the observation of a very
compact (or what seems to be a very compact) object revives the question whether
strange stars can actually exist. Thus, it is interesting to see whether
gravitational waves would be able to provide an answer to this question. 
For this reason in \cite{strangestars} we computed the $f$-mode
eigenfrequency of strange stars, modeled using the MIT bag model \cite{bagmodel}.
It may be noted that, due to the complexity of the fundamental theory of strong interactions
between quarks (Quantum Chromo-Dynamics, or QCD), theoretical studies
of strange stars are necessarily based on models, and the MIT bag model is one of
the most used in the literature.
According to such model, quarks occur in color neutral
clusters confined to a finite region of space -- the bag -- the volume of
which is limited by the pressure of the QCD vacuum (the bag constant
$B$);  in addition, the residual interactions between quarks are assumed 
to be weak, and therefore are treated in low order perturbation 
theory in the color coupling constant $\alpha_s$.
Thus, the parameters of the model are the masses of up, down and strange
quarks,  $\alpha_s$ and the bag constant $B$. 

From the Particle Data Book we learn that the mass of the up and down quarks
are of the order of few MeV,  negligible with respect to that of
the strange quark, the value of which is in the ranges  $(80 -155)$~  MeV.
The value of the coupling constant $\alpha_s$ is constrained by the results of
hadron collision experiments to range within $(0.4 -0.6)$.

\begin{figure}[ht]
\begin{center}
\includegraphics[width=7cm,angle=270]{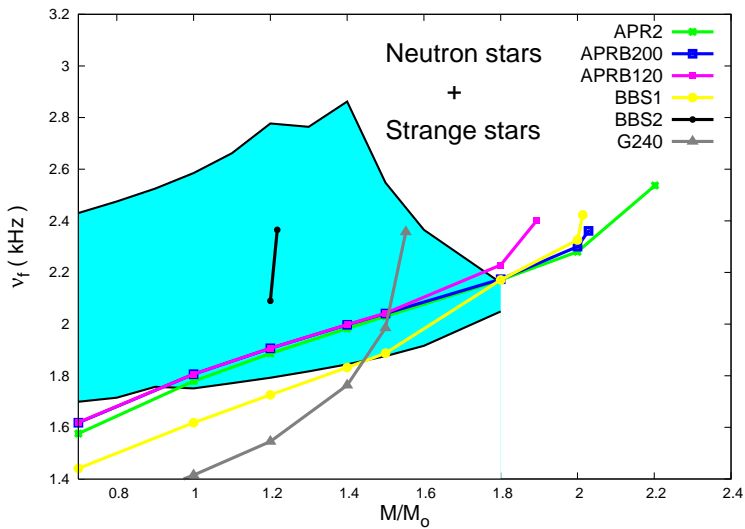}
\caption{The frequency of the fundamental mode is
plotted as a function of the mass of the star, for neutron/hybrid stars
(continuous lines)
and for strange stars modeled using the MIT bag model, spanning the set of
parameters indicated in (\ref{param}) (dashed region).}
\label{SS}
\end{center}
\end{figure}

In early applications of the MIT bag model $B$, $\alpha_s$ and $m_s$ were
adjusted to fit the measured properties of light hadrons (spectra,
magnetic moments and charge radii).
According to these studies $B$ was shown to range 
from $57.5$ MeV/fm$^3$ \cite{DeGrand75} to $351.7$ MeV/fm$^3$
\cite{Carlson83}; however,
the requirement that strange quark matter be absolutely stable at zero
temperature and pressure implies that $B$ cannot exceed the maximum
value $B_{max}\,\approx 95$\,MeV/fm$^3$~\cite{Farhi84}.  For
values of $B$ exceeding $B_{max}$, a star entirely made of deconfined
quarks is not stable,  and  quark matter can only
occupy a fraction of the available volume
 as in the models \apba and \apbb considered above.
Thus, if we want to study bare strange stars we need to restrict the values of $B$
in the range $\in (57 -95)$\,MeV/fm$^3$.
In our analysis we have systematically explored the following range of parameters 
\be
m_s \in (80 -155)~ {\rm MeV},\qquad 
\alpha_s \in (0.4 -0.6),\qquad B \in (57 -95)\,{\rm MeV}/{\rm fm}^3,
\label{param}
\ee
computing the corresponding stellar configurations up to the maximum mass, and
the corresponding $f$-mode frequencies. The results are summarized in figure
\ref{SS}. There we plot $\nu_f$
as a function of the mass of the star, both for strange stars and
for the neutron/hybrid stars described in the previous section.
The shaded region covers the range of parameters of the MIT
bag model  (\ref{param}).

From figure \ref{SS} we can extract the following information.
First of all strange stars cannot emit gravitational waves
with $\nu_f \lesssim 1.7$ kHz, for any value of the mass in the range
we consider.
Note that $1.8~M_\odot$ is the maximum mass 
above which no stable strange star can exist. 
There is a small range of frequency where neutron/hybrid stars are
indistinguishable from strange stars; however, there is a large
frequency region where only strange stars can emit. For instance if
$M=1.4~ M_\odot$, a signal with $\nu_f \gtrsim 2$~kHz would belong
to a strange star.
Even if we do not know the mass of the star
(as it is often the case for isolated pulsars) the knowledge of
$\nu_f$ allows to gain information about the source nature; indeed, if
$\nu_f \gtrsim 2.2$ kHz, apart from a very narrow region of masses where
stars with hyperons would emit (EOS \BBSB and \GG),
we can reasonably exclude that the signal is
emitted by a neutron star.

In addition, it is possible to show that if a signal emitted by an oscillating strange star
would be detected, since $\nu_f$ is an increasing function of the bag constant $B$
it would be possible to set constraints on $B$ much more stringent than those provided by the
available experimental data \cite{strangestars}.

\section{Are we ready for gravitational wave asteroseismology?}
In this section we want to discuss  whether we are in  a position to
establish what is (or are) the equation(s) of state appropriate 
to describe  matter at supranuclear densities, using gravitational wave signals.
This is of course a fundamental question, because the energies prevailing in
the inner core of  neutron stars are unaccessible to high energy experiments on
Earth.

The answer is, unfortunately, negative,
essentially for two reasons. The first is that ground based interferometers do
not have, at present,  sufficiently high
sensitivity  at  frequencies in
the range  $\sim 1.5 - 3$ kHz, typical for $\nu_f$. Feasibility studies of interferometric, 
high frequency detectors have been considered in recent years \cite{euro},
and high frequency, wide-band,  resonant detectors are under study \cite{dual};
however, if we restrict to  Virgo or LIGO in their present
configuration,  to detect a signal emitted by a NS pulsating in the $f$-mode in our
Galaxy, with a signal-to-noise ratio of 5,
the energy stored into the mode
should be $E_{f-mode} \sim 6\cdot 10^{-7}~M_\odot c^2$.
In order to understand whether it is plausible that the fundamental mode is  excited
to such an extent, we can refer either to the results of numerical simulations, or to
astrophysical data. Numerical simulations of gravitational collapse 
show that the amount of energy 
released in gravitational waves is in the range
$E_{GW~tot}\sim [10^{-9}-10^{-6}]~M_\odot c^2$ \cite{collapse}. 
Although  computed waveforms show that the $f$-mode is excited,
at present there is no conclusive indication on the
fraction of $E_{GW~tot}$ which may go into that mode, since
it depends on the initial conditions and on the physical assumptions that are 
made in modeling the collapse. Just to mention one, usually
numerical simulations assume axisymmetric collapse, but in the non axisymmetric case
energy released in GWs may be higher.

Thus, we can only say that $E_{f-mode} \sim 6\cdot
10^{-7}~M_\odot c^2$ is not totally unreasonable.  Unfortunately,
gravitational collapse is a rare event (about 3 events every hundred years, per galaxy), 
and if we restrict to our Galaxy chances to detect one in our lifetime are not too high.

The $f$-mode may also be excited in a cold, old neutron star
as a consequence of a glitch.
Glitches are sudden changes in the rotation frequency of the
neutron star crust.  They are observed in many pulsars and are thought to be
related to quakes occurring in the solid structures such as the crust, the
superfluid vortices and, perhaps, the lattice of quark matter in the
stellar core \cite{gl1,gl2,gl3}.  The rotational energy released in a glitch is
$~\Delta E \approx I\Omega\Delta\Omega$, where $I$ is the
moment of inertia of the star, and
typical spin variations are  $\Delta\Omega /\Omega
\approx 10^{-6} - 10^{-8}$. 
For the glitches observed in the Crab and Vela pulsars
observations give $~\Delta E \simeq 2\cdot 10^{-13}~M_\odot c^2$ and $~\Delta E \simeq 3\cdot
10^{-12}~M_\odot c^2$, respectively. As before, we do not know which fraction of $\Delta E$
goes in the $f$-mode excitation; in any event
being $\Delta E$ so small, we can conclude that
there is no hope to detect anything like this with the detectors that are actually
in operation.

\begin{figure}[ht]
\begin{center}
\includegraphics[width=7cm,angle=270]{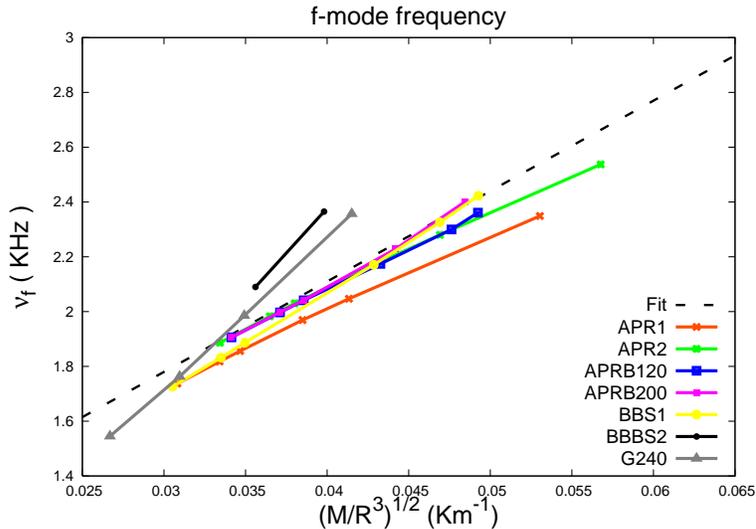}
\caption{The frequency of the fundamental mode is plotted 
as a function of the square root of the average density for the 
EOS considered in this paper.
We also plot the fit given in eq. (\ref{fitf}) }
\label{nuf}
\end{center}
\end{figure}

The second reason why we are far from being able to infer the EOS of matter
in the inner core of a NS using gravitational waves, is that the EOS proposed in the literature
only loosely constrain the dynamics of  nuclear matter. 
This statement can be made more clear by the following example. 
Since from Newtonian gravity we know that  $\nu_f$ scales as the
square root of the average density, we expect a similar relation to hold 
also in general relativity. Indeed, a linear relation 
between $\nu_f$ and $\sqrt{\frac{M}{R^3}}$ has
been obtained in \cite{AK}, fitting the data referring to stars modeled with a
large set of EOS.  In \cite{astero_nostro} 
a similar fit has been found using the same EOS
considered in previous sections; since the two fits are similar, 
in what follows we shall explicitly use our fit:
\be
\nu_f=a+b\sqrt{\frac{M}{R^3}},\quad
a=0.79\pm0.09,\quad b=33\pm2 ,
\label{fitf}
\ee
where $a$ is in kHz and $b$ in km$\cdot$kHz.
The fit is plotted in figure \ref{nuf} versus $\sqrt{\frac{M}{R^3}}$,
together with the $f$-mode frequencies 
corresponding to the stellar models considered in figure \ref{nufM}.
Similarly,  the first $p$-mode frequency
can be fitted as a function of the stellar compactness $M/R$
as follows
\be
\nu_{p_1}=\frac{1}{M}\left[a+b\frac{M}{R}\right],\qquad
a=-1.5\pm0.8,\quad b=79\pm4,
\label{fitp}
\ee
where $a$  and $b$ are in km$\cdot$kHz.
In these fits, 
frequencies are expressed in kHz, masses and radii in km.

\begin{figure}[ht]
\begin{center}
\includegraphics[width=7cm,angle=270]{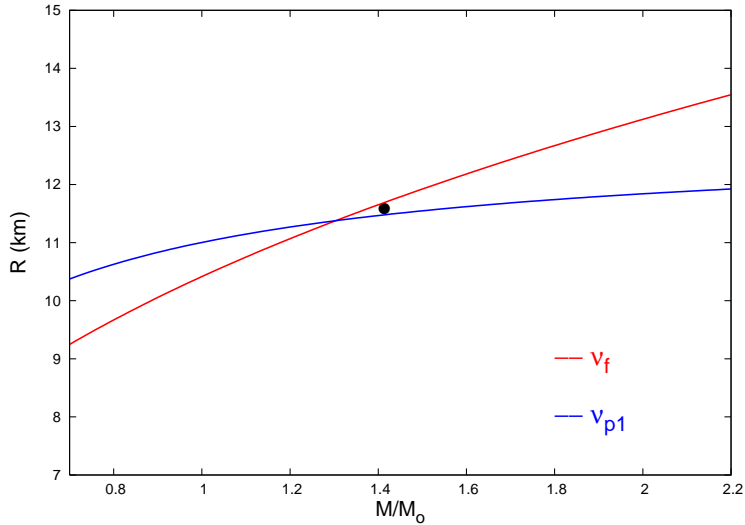}
\caption{The fits (\ref{fitf}) and (\ref{fitp}) are plotted in the $R-M$-plane,
assuming the mode frequencies $\nu_f=1.983$~kHz and $\nu_p=6.164$~kHz have been identified 
in a detected signal. The black dot corresponds to the true values of the mass and radius of
the emitting star.}
\label{FIT_ideal}
\end{center}
\end{figure}

\begin{figure}[ht]
\begin{center}
\includegraphics[width=7cm,angle=270]{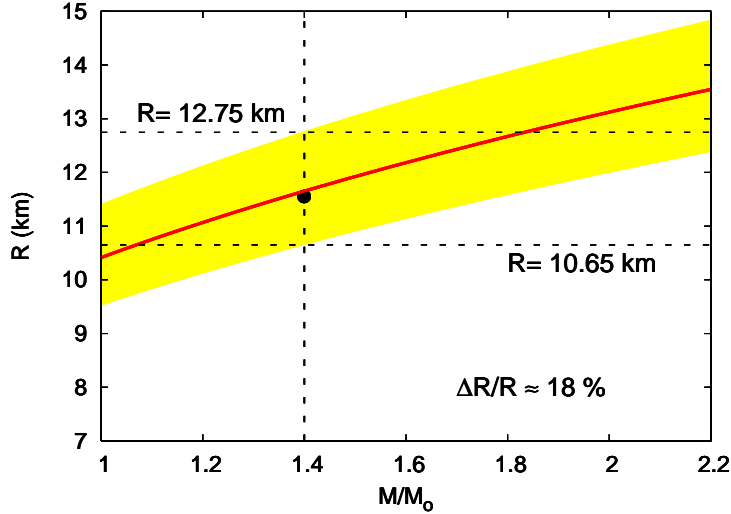}
\caption{The fit (\ref{fitf}) is plotted in the $R-M$-plane including the error
bars on the fit parameters (dashed region).}
\label{fitraggio}
\end{center}
\end{figure}

Let us now consider  a star belonging to the EOS \aprb, with mass
$M=1.4~M_\odot$ and radius $R=11.58$~km. 
Let us assume that, as a consequence of some astrophysical
process, both the fundamental mode and  the first $p$-mode are excited  
and that the emitted gravitational wave has been detected.
With the detected values of $\nu_f$ and $\nu_{p_1}$ 
(which we know to be $\nu_f=1.983$~kHz, $\nu_p=6.164$~kHz)
we could plot the fits
(\ref{fitf}) and (\ref{fitp}) in the
$(R,M)$-plane, and we would find what is shown in figure \ref{FIT_ideal}: the two
curves intersect in a point which corresponds to  $M=1.30~M_\odot, ~R= 11.36$~km; 
consequently, we would be able to estimate
the values of the mass and of the radius with an error of $7\%$ and $2\%$, respectively.
This would be great, but unfortunately, the fit comes with  error bars. 
If, for instance we
plot the curve referring to the $f$-mode, and we show the entire region where the
parameters $a$ and $b$ can vary (the dashed region in figure \ref{fitraggio}), 
we see that the error bar induces a very large
error on $R$;
so large indeed that, even knowing the mass, we would estimate 
$R$ with an error of the order of  $18\%$. 

Therefore, for the time being, we can only say that  gravitational wave asteroseismology 
will become possible
when GW-detectors will become  more sensitive to the
high frequency region,
and when nuclear matter studies will put tighter constraints on  the 
parameters that characterize the equation of state  of superdense matter.

Since science always looks forward for expanding the horizon of knowledge,
we are confident that one day this will be possible.


\begin{thebibliography}{99}
\bibitem{webaddress} {\tt www.ligo.caltech.edu}; {\tt www.virgo.infn.it}
\bibitem{Kostasrev} Kokkotas K.D. and Schmidt B.G. 1999 {\it Living Rev. Rel.} 
{\bf 2}, 2 
\bibitem{Nollertrev} Nollert H.P. 1999 {\it Class. Quant. Grav.} 
{\bf 16} R159
\bibitem{sha} Shapley H. 1914 {\it Astrophys. J.} {\bf 40} 448
\bibitem{edd1} Eddington A.S. 1918 {\it Mon Not Roy. Astron. Soc.}
{\bf 79}, 2
\bibitem{regwe} Regge T. and Wheeler J.A. 1957 {\it Phys. Rev.} 
{\bf 108}, 1063 
\bibitem{zer}
Zerilli J.F. 1970 {\it Phys. Rev.} D {\bf 2}, 2141;
Zerilli J.F. 1970 {\it Phys. Rev. Lett.} {\bf 24}, 737
\bibitem{MTW} Misner C.W., Thorne K S and Wheeler J A 1973,
{\it Gravitation}, W.H. Freeman \& C, New York
\bibitem{vish} Vishveshwara C.V. 1970 {\it Phys. Rev.} D 
{\bf 1}, 2870 
\bibitem{pr1} Press W.H. 1971 {\it Astrophys. J.} {\bf 170}, L105
\bibitem{chdet}
Chandrasekhar S. and Detweiler S.L. 1975 {\it Proc. R. Soc. Lond.} 
A {\bf 344}, 441 
\bibitem{mt}
Chandrasekhar S. 1984 {\it The mathematical theory of black hole},
Claredon Press, Oxford
\bibitem{Ghez:2003qj}
Ghez A.M., Salim S., Hornstein S.D., Tanner A., Morris M.,
Becklin E.E. and Duchene G. 2005, {\it Astrophys. J.} {\bf 620}, 744 
\bibitem{wkb1}
Schutz B.F. and Will C.M. 1985 {\it Astrophys. J. Lett.} {\bf 291}, L33;
Iyer S. and Will C.M. 1987 {\it  Phys. Rev.}, D {\bf 35}, 3621;
Iyer S. 1987 {\it  Phys. Rev.} D {\bf 35}, 3632 (1987);
Kokkotas K.D. and Schutz B.F. 1988 {\it  Phys. Rev.} D {\bf 37}, 12
\bibitem{wkb2} 
Guinn J.W., Will C.M., Kojima Y. and Schutz B.F. 1990
{\it Class. Quantum Grav.} {\bf 7}, L47 
\bibitem{leaver}
Leaver E.W. 1985 {\it Proc. R. Soc. Lond.} A {\bf 402}, 285
\bibitem{phase} Andersson N.,1992 {\it Proc. R. Soc. Lond.} A 
{\bf 439}, 47
\bibitem{Asint} Nollert H.-P. 1993 {\it Phys. Rev.} D {\bf 47}, 5253;
Andersson N. 1993 {\it  Class. Quantum Grav.} {\bf 10}, L61; 
Barreto A.S. and Zworski M. 1997 {\it Math. Res. Lett.} {\bf 4}, 103;
Padmanabhan T. 2004 {\bf Class. Quantum Grav.} 21, L1;
Motl L. 2003 {\it Adv. Theor. Math. Phys.} {\bf 6}, 1135 
\bibitem{Fiziev} Fiziev P.P. 2006 {\it Class. Quantum Grav.} {\bf 23},
2447
\bibitem{teuk}
Teukolsky S. 1972 {\it Phys. Rev. Lett.} {\bf 29}, 1114;
Teukolsky S. 1973 {\it Astrophys. J.} {\bf 185}, 635
\bibitem{starchu}
Starobinski A.A. and Churilov S.M. 1973 {\it Soviet JEPT} {\bf 38}, 1
\bibitem{prteu}
Press W.H. and  Teukolsky S. 1973 {\it Astrophys. J.} {\bf 185}, 649 
\bibitem{detkerr}
Detweiler S.L. 1977 {\it Proc. R. Soc. Lond.} A {\bf 352}, 381;
Detweiler S.L. 1979 in {\it Sources of Gravitational Radiation}, edited by
L. Smarr, Cambridge, England, 211;
Detweiler S.L. 1978 {\it Astrophys. J.} {\bf 225}, 687;
Detweiler S.L. 1980 {\it Astrophys. J.} {\bf 239}, 292 
\bibitem{seiy}
Seidel E., Iyer S. 1990 {\it Phys. Rev} D {\bf 41}, 374
\bibitem{kk1}
Kokkotas K.D. 1991 {\it Class.Quantum Grav.} {\bf 8}, 2217
\bibitem{recentKerrmodes} Onozawa H. 1997 {\it Phys. Rev.} D {\bf 55}, 3593   
\bibitem{mashoonferrari}
Ferrari V. and Mashoon B. 1984 {\it Phys. Rev. Lett.} {\bf 52}, 1361;
Ferrari V. and Mashoon B. 1984 {\it Phys. Rev.} D {\bf 30}, 295
\bibitem{AsintKerr} Berti E., Cardoso V., Kokkotas K.D. and
Onozawa H. 2003 {\it Phys. Rev.} D {\bf 68}, 124018;
Berti E., Cardoso V. and Yoshida S. 2004 {\it Phys. Rev.} D {\bf 69}, 124018 
\bibitem{DRPP} Davis M., Ruffini R., Press W.H. and Price R.H. 1971 
{\it Phys. Rev. Lett} 27, 1466
\bibitem{RTV} Davis M., Ruffini R. and Tiommo J. 1971 
{\it Phys. Rev.} D {\bf 5}, 2932; Ferrari V. and Ruffini R. 1981 
{\it Phys. Lett.} B {\bf 98}, 381  
\bibitem{SW} Haugan M.P., Shapiro S.L. and Wasserman I. 1982 
{\it Astrophys. J.} {\bf 257} 283;
Shapiro S.L. and Wasserman I. 1982 {\it Astrophys. J.} 
{\bf 260} 838;
Pertich L.I., Shapiro S.L. and Wasserman I. 1985 {\it Astrophys. 
J. Suppl. S} {\bf 58} 297
\bibitem{ON} Oohara K. and Nakamura T. 1983 {\it Phys. Lett.} A
{\bf 98} 403
\bibitem{BCW} Berti E., Cardoso V. and Will C.M. 2006
{\it AIP Conf. Proc.} {\bf 848}, 687 
\bibitem{NOK} Nakamura T., Oohara K. and Kojima Y. 1987 
{\it Prog. Theor. Phys. Suppl.} {\bf 90} 218  
\bibitem{numcoal} Pretorius F. 2005 {\it Phys. Rev. Lett} {\bf 95} 
121101; 
Campanelli M., Lousto C.O., Marronetti P. and Zlochower Y., 2006
{\it Phys. Rev. Lett} {\bf 96} 111101; 
Baker J.G., Centrella J., Choi D.-I., Koppitz M. and van Meter J. 2006
{\it Phys. Rev. Lett} {\bf 96} 111102
\bibitem{comparcoal} Baker J.G., Campanelli M., Pretorius F. and 
Zlochower Y. 2007 {\it Class. Quant. Grav.} {\bf 24}, S25 
\bibitem{coalkerr} Campanelli M., Lousto C.O. and Zlochower Y. 2006
{\it Phys. Rev.} D {\bf 74}, 041501
\bibitem{BHRS} Baiotti L., Hawke I., Rezzolla L. and Schnetter E. 2005
{\it Phys. Rev. Lett.}  {\bf 94}, 131101
\bibitem{ZFR} Zanotti O., Font J.A., Rezzolla L. and Montero P.J. 2005
{\it Mon Not Roy. Astron. Soc.}  {\bf 356}, 1371;
Nagar A., Zanotti O., Font J.A. and Rezzolla L. 2007 
{\it Phys. Rev.} D {\bf 75}, 044016 
\bibitem{FGR} Ferrari V., Gualtieri L. and Rezzolla L. 2006
{\it Phys. Rev.} D {\bf 73}, 124028 
\bibitem{leaver2} Leaver H.W. 1986 {\it Phys. Rev.} D {\bf 34}, 384
\bibitem{residues} 
Sun Y., and Price R. H. 1988 {\it Phys. Rev.} D {\bf 38}, 1040;
Andersson, N. 1995 {\it Phys. Rev.} D {\bf 51}, 353;
Nollert P. and Price R.H. 1999 {\it J. Math. Phys.}  {\bf 40}, 980;
Glampedakis K. and Andersson N. 2001 {\it Phys. Rev.} D {\bf 64}, 104021;
Berti E. and Cardoso V. 2006 {\it Phys. Rev.} D {\bf 74}, 104020 
\bibitem{price72} Price R.H. 1972 {\it Phys. Rev.} D {\bf 5} 2419
\bibitem{Vishstab} Vishveshwara C.V. 1970 {\it Phys. Rev.} D {\bf 1}, 2870
\bibitem{Waldstab} Wald R.M. 1979 {\it J. Math. Phys.} {\bf 20}, 1056;
Kay B.S. and Wald R.M. 1987 {\it Class. Quantum Grav.} {\bf 4}, 893 
\bibitem{KerrStability} Whiting B.F. 1989 {\it J. Math. Phys.} 
{\bf 30}, 1301; Beyer H.R. 2001 {\it Comm. Math. Phys.} {\bf221}, 659 
\bibitem{HH} Horowitz G.T. and Hubeny V.E. 2000 {\it Phys. Rev.}
D {\bf 62}, 024027
\bibitem{AdS} Wang B., Lin C.Y. and Abdalla E. 2000 {\it Phys. Lett.}
B {\bf 481}, 79;
Wang B., Molina C. and Abdalla E. 2001 {\it Phys. Rev.} D {\bf 63}, 084001;
Cardoso V. and Lemos J.P.S. 2001 {\it Phys. Rev.} D {\bf 63}, 124015;
Cardoso V. and Lemos J.P.S. 2001 {|IT Phys. Rev.} D {\bf 64}, 084017;
Berti E. and Kokkotas K.D. 2003 {\it Phys. Rev.} D {\bf 67}, 064020;
Cardoso V., Konoplya R. and Lemos J.P.S. 2003 {\it Phys. Rev.} D {\bf 68},
044024
\bibitem{AdSold} Chan J.S.F. and Mann R.B. 1996 {\it Phys. Rev.} D 
{\bf 55}, 7546; 1999 {\it Phys. Rev.} D {\bf 59}, 064025.
\bibitem{DM} Dreyer O. 2003 {\it Phys. Rev. Lett.} {\bf 90}, 081301;
Motl L. 2003 {\it Adv. Theor. Math. Phys.} {\bf 6}, 1135
\bibitem{Asint2} Motl L. and Neitzke A. 2003 {\it Adv. Theor. Math. Phys.}
{\bf 7}, 307;
Cardoso V., Natario J. and Schiappa R. 2004 {\it J. Math. Phys.}
{\bf 45}, 4698;
Natario J. and Schiappa R. 2004 {\it Adv. Theor. Math. Phys.} {\bf 8},
1001.
\bibitem{VariousQNM} Cardoso V. and Lemos J.P.S. 2003 {\it Phys. Rev.}
D {\bf 67}, 084020;
Konoplya R.A. 2003 {\it Phys. Rev.} D {\bf 68}, 024018;
Cardoso V., Lemos J.P.S. and Yoshida S. 2004 {\it Phys. Rev.}
D {\bf 69}, 044004
\bibitem{Nick} Stergioulas N. 2003 {\it Living Rev. Rel.} {\bf 6}, 3 
\bibitem{TC} Thorne K.S. and Campolattaro A. 1968 {\it Astrophys. J.}
{\bf 149}, 591; 1970 {\it Astrophys. J.} {\bf 159}, 847
\bibitem{CF} Chandrasekhar S. and Ferrari V. 1991 {\it Proc. R. Soc. Lond.}
A {\bf 432}, 247;
\bibitem{VF} Ferrari V. 1992 {\it Phil. Trans. R. Soc. Lond.} A {\bf 340}, 423
\bibitem{D} Detweiler S.L. 1975 {\it Astrophys. J.} {\bf 201}, 440
\bibitem{LD} Lindblom L. and Detweiler S.L. 1983 {\it Astrophys. J.
Suppl.} {\bf 53}, 73
\bibitem{LS} Lindblom L. and Splinter R.S. 1989 {\it Astrophys. J.}
{\bf 345}, 925
\bibitem{DL} Detweiler S.L. and Lindblom L. 1985 {\it Astrophys. J.}
{\bf 292}, 12
\bibitem{ipserprice} Ipser J.R. and Price R.H. 1992 
{\it Phys. Rev.} D {\bf 43} 1768
\bibitem{wkostas} Kokkotas K.D. 1994 {\it Mon. Not. R. Ast. Soc.} 
{\bf 268} 1015
\bibitem{axialmodes}
Chandrasekhar S. and Ferrari V. 1991 {\it Proc. R. Soc. Lond.}
{\bf 434}, 449
\bibitem{omarema} Benhar O., Berti E. and Ferrari V. 1999
{\it Mon. Not. Roy. Astron. Soc.}  {\bf 310}, 797
\bibitem{cowling} Cowling T.G. 1942 {\it Mon. Not. R. Ast. Soc.}
{\bf 101}, 367
\bibitem{kokkoschutzwmodes}
Kokkotas K.D. and Schutz B.F. 1992 {\it Mon. Not. R. Ast. Soc.} {\bf 255} 119
\bibitem{AK} Andersson N. and Kokkotas K.D. 1998 
{\it Mon. Not. R. Ast. Soc.} {\bf 299} 1059
\bibitem{astero_nostro} Benhar O., Ferrari V. and  Gualtieri L. 2004
{\it Phys. Rev.} D {\bf 70}, 124015
\bibitem{BPS}
Baym G., Pethick C.J. and Sutherland P. 1971 
{\it Astrophys. J.} {\bf 170}, 299
\bibitem{PRL} Pethick C.J., Ravenhall B.G. and Lorenz C.P. 1995 
{\it Nucl. Phys.} A {\bf 584}, 675
\bibitem{AP}
Akmal A. and Pandharipande V.R. 1997 {\it Phys. Rev.} C {\bf 56}, 2261
\bibitem{APR}
Akmal A., Pandharipande V.R. and Ravenhall D.G. 1998 {\it Phys. Rev.} 
C {\bf 58}, 1804
\bibitem{BBS200} Baldo M., Burgio G.F. and Schulze H.J. 2000
{\it Phys. Rev.} C, {\bf 61}, 055801
\bibitem{BR}
Rubino R., thesis, Universit\`a ``La Sapienza", Roma (unpublished);
Benhar O. and Rubino R., to be published.
\bibitem{Gbook}
Glendenning N.K. 2000 {\it Compact Stars} (Springer, New York)
\bibitem{lejeune}
Lejeune A., Grang\'e P., Martzoff P. and Cugnon J. 1986
{\it Nucl. Phys.} A {\bf 453}, 189
\bibitem{Witten}
Witten E. 1984 {\it Phys. Rev.} D {\bf 30}, 272
\bibitem{strangestars} Benhar O., Ferrari V., Gualtieri L. 
and Marassi S. 2007 {\it Gen. Rel. Grav.} {\bf 39}, 1323
\bibitem{bagmodel}
Chodos A., Jaffe R.L., Johnson K., Thorne C.B. and Weiskopf W.F. 1974
{\it Phys. Rev.} D {\bf9},3471
\bibitem{DeGrand75}
De Grand T., Jaffe R.L., Johnsson K. and Kiskis J. 1975
{\it Phys. Rev.} D {\bf12}, 2060
\bibitem{Carlson83}
Carlson C.E., Hansson T.H. and Peterson C. 1983 {\it Phys. Rev.} D 
{\bf27}, 1556
\bibitem{Farhi84}
Farhi E. and Jaffe R.L. 1984 {\it Phys. Rev.} D {\bf 30}, 2379
\bibitem{euro} {\tt www.astro.cf.ac.uk/geo/euro}
\bibitem{dual} {\tt www.dual.lnl.infn.it}
\bibitem{collapse}
Dimmelmeier H., Font J.A. and M\"uller E. 2002
{\it Astron. Astrophys.} {\bf 393}, 523;\\
M\"uller E., Rampp M., Buras R., Janka H.-T. and
Shoemaker D.H. 2004 {\it Astrophys. J.} {\bf 603}, 221;\\
Ott C.D., Burrows A., Dessart L. and Livne E. 2006
{\it Phys. Rev. Lett.} {\bf 96}, 201102.

\bibitem{gl1} Anderson P.W. and Itoh N. 1975, {\it Nature} {\bf 256}, 25
\bibitem{gl2} Ruderman M. 1976, {\it Nature} {\bf 203}, 213
\bibitem{gl3} Pines D. and Alpar M.A. 1985, {\it Nature} {\bf 316}, 27
\end{thebibliography}
\end{document}